\documentclass[runningheads]{llncs}

\usepackage[colorlinks=true]{hyperref}
\usepackage{longtable}
\usepackage{listings}
\usepackage{amsmath}
\usepackage{csquotes}
\usepackage{cite}
\usepackage[T1]{fontenc}
\usepackage{graphicx}
\usepackage[misc]{ifsym}
\usepackage{orcidlink}

\newcommand{\tinyskip}{\hspace{0.7em}}
\newcommand{\expr}{\textsc{expr}}
\newcommand{\stmt}{\textsc{stmt}}
\newcommand{\type}{\textsc{type}}
\newcommand{\dest}[1]{#1\texttt{\_dest}}

\def\go{\lstinline}
\begin{document}

\title{Extending Isabelle/HOL's Code Generator with support for the Go programming language}

\titlerunning{Go support for the Isabelle Code Generator}

\author{Terru Stübinger\inst{1,2}\orcidlink{0009-0006-7411-2533} \and
Lars Hupel\inst{1,2}\orcidlink{0000-0002-8442-856X}}

\authorrunning{T. Stübinger, L. Hupel}

\institute{Giesecke+Devrient, Prinzregentenstr. 161, 81677 München, Germany \and
Technische Universität München, School of Computation, Information and Technology, Boltzmannstr. 3, 85748 Garching bei München, Germany\\
\email{stuebinm@in.tum.de, lars.hupel@tum.de}}

\maketitle

\begin{abstract}
The Isabelle proof assistant includes a small functional language, which allows users to write and reason about programs.
So far, these programs could be extracted into a number of functional languages: Standard ML, OCaml, Scala, and Haskell.
This work adds support for Go as a fifth target language for the Code Generator.
Unlike the previous targets, Go is not a functional language and encourages code in an imperative style,
thus many of the features of Isabelle's language (particularly data types, pattern matching, and type classes) have to be emulated using imperative language constructs in Go.
The developed Code Generation is provided as an add-on library that can be simply imported into existing theories.

\end{abstract}

\section{Introduction}

The interactive theorem prover \emph{Isabelle} of the LCF tradition~%
\cite{concrete-semantics} is based on a small, well-established and trusted mathematical inference kernel written
in Standard ML.\@ All higher-level tools and proofs, such as those included in the most
commonly-used logic \emph{Isabelle/HOL}, have to work through this kernel.

Many of the tools available to users in \emph{Isabelle/HOL}
feel immediately familiar to anyone with experience in functional programming
languages: it is possible to define data types,
functions, and Haskell-style type classes and instances.

Isabelle's nature as a theorem prover further makes it easy
to formalise and prove propositions about such programs.
To allow use of such programs outside of the proof assistant's environment,
Isabelle comes equipped with a \emph{Code Generator}, allowing users to
extract source code in Haskell, Standard ML, Scala, or OCaml, which can then
be compiled and executed. This translation of code works by first translating into
an intermediate language called \emph{Thingol}, shared between all targets;
from this language, code is then transformed into the individual target
languages via the principle of \emph{shallow embedding}, that is, by representing
constructs of the source language using only a well-defined subset of the target
language, thus side-stepping the issue of finding a complete formal description of
a target language's behaviour~\cite{haftmann-phd,haftmann-nipkow}.

\emph{Go} is a programming language introduced by Google in 2009.
It is a general-purpose, garbage-collected, and statically
typed language \cite{gospec}. In contrast to the existing targets of Isabelle's
Code Generator, it is not a functional language, and encourages programming in
an imperative style.
However, it is a very popular language, and
many large existing code bases have been written in it.

\paragraph{Contributions}

This paper extends Isabelle's Code Generation facility with support for Go.
For that, we demonstrate a translation scheme from programs in Thingol
to programs in Go (§\ref{sec:translation}).
We provide this facility as a stand-alone theory file that
can easily be imported into existing developments.
We provide our development as an entry in the \emph{Archive of Formal Proofs (AFP)}---a repository of Isabelle proof libraries---,
making it immediately usable in other contexts \cite{Go-AFP}.

The motivation for this work stems from the internal use of both ecosystems at Giesecke+Devrient:
Isabelle for formalisation purposes, and Go for the real-world implementation.
This naturally lead to a formalisation gap, which this project sought to close
(§\ref{sec:evaluation}).

\paragraph{Related work}

This paper describes the first attempt at translating Isabelle formalisations
into a non-functional programming language.
Prior work in leveraging imperative features in the Code Generator~\cite{imperative-hol} has targeted the existing, functional programming languages, and thereby could reuse much of the existing infrastructure.
There is also unpublished work on adding support for F\# to the Code Generator~\cite{brucker-fsharp}, another functional language.

Shallow embeddings of C in proof assistant are already well known; for example in F* \cite{protzenko-fstar}, Isabelle \cite{Simpl-AFP}, and Why3 \cite{rieu}.
Those tools are not designed to export arbitrary code, but require developers to use a restricted subset of the host language.
Instead, they are mainly geared towards low-level programming; with some providing C-style memory management.
Our work focuses instead on translating the full functional host language into a high-level imperative language, therefore avoiding the need to (re)write host language code specifically for the purpose.

\section{The intermediate language Thingol}\label{sec:thingol}

Isabelle's Code Generation pipeline works in multiple stages. Crucially,
all definitions made in Isabelle are first translated into an abstract
intermediate language called \emph{Thingol}, which is the last step shared
between all target languages. The final stage then uses a shallow embedding to
translate the Thingol program into source code of the target language.

Consequently, Thingol's design reflects the features common to previous
target languages, and is based on a simply-typed $\lambda$-calculus with ML-style
polymorphism. Perhaps surprisingly, Thingol also supports type classes, which
can be mapped easily to Haskell and Scala, but less easily to the other targets,
which instead use a dictionary construction (§\ref{sec:translation:dictionaries}).
The supported fragment of type classes and instance corresponds to Haskell98,
with the exception of constructor classes (which would require a more expressive type system) \cite{haftmann-classes,jones2003haskell}.

\begin{figure}[t]
\begin{lstlisting}[basicstyle=\ttfamily\small]
datatype Nat = Zero | Suc Nat
datatype $\alpha$ list = Nil | Cons $\alpha$ ($\alpha$ list)

fun hd2 :: $\forall \alpha. \alpha\;\mathtt{list} \Rightarrow \alpha\;\mathtt{option}$ where
  hd2 $xs$ = case $xs$ of Nil $\Rightarrow$ None
                      | Cons $x$ Nil $\Rightarrow$ None
                      | Cons $x$ (Cons $y$ $\mathit{xs}$) $\Rightarrow$ Some $y$

class semigroup where
  (+) :: $\alpha \Rightarrow \alpha \Rightarrow \alpha$

class monoid $\subseteq$ semigroup where
  zero :: $\alpha$

instance Nat :: semigroup where
  $\mathit{a}$ + Zero = $\mathit{a}$
  Zero + $\mathit{a}$ =  $\mathit{a}$
  (Suc $\mathit{a}$) + $\mathit{b}$ = Suc ($\mathit{a}$ + $\mathit{b}$)

instance Nat :: monoid where
  zero = Zero

fun sum :: ($\alpha$ :: monoid) list $\Rightarrow$ $\alpha$ where
  sum $\mathit{xs}$ = fold (+) $\mathit{xs}$ zero
\end{lstlisting}
  \caption{An example program (omitting the definition of \lstinline|fold| for brevity)}
\label{listing:thingol}
\end{figure}

Thingol's terms are simple $\lambda$-expressions with the addition of case expression for pattern matching on data types.
A Thingol program is a list of declarations, i.e.\ top-level items which introduce data types, functions, type classes, and their instances.

While there is no formal semantics of Thingol, it can be thought of as a \emph{Higher-Order Rewrite System} (HRS)~\cite{nipkow-hrs,mayr-hrs}.
It provides a convenient abstraction over the target languages' semantics.
Because a HRS does not have a specified evaluation order, the Code Generator cannot guarantee total, but only partial correctness.
(This restriction applies to all supported target languages.)

Reusing Thingol has two immediate benefits: we can leverage the entire entire
existing pipeline as well as its existing code adaptations, and are not forced to reimplement some tedious translation of Isabelle's more advanced features.
Additionally, creating a custom intermediate language would not help to meaningfully address the functional--imperative mismatch between Isabelle/HOL and Go, but only shift the complexity elsewhere.

\section{The target fragment of Go}\label{sec:go}

Go, being an imperative language, differs in many aspects from the already-existing target languages
of Isabelle's Code Generator.
Conversely, many of Go's unique features are not needed by the generator. Since
the translation works as a shallow embedding into the target language, it suffices
to use the fragment which can be used to represent the various statements
of Thingol. Consequently, we will focus on this fragment only, but discuss---if
necessary---why we did not pursue alternative features or solutions.

This approach leaves many of Go's most interesting features (e.g.\ channels or methods) entirely unused.
The fragment we use can be understood as a
\enquote{functional subset} of the Go language, meaning that it comprises only those
features that closely align with those of the (functional) pre-existing code
generation targets available in Isabelle as well as those of Thingol.

\subsection{Syntax}\label{sec:go:syntax}

The syntactic fragment used by the Code Generator%
\footnote{See the appendix of this paper for the full listing: \href{https://arxiv.org/abs/2310.02704}{\texttt{arxiv.org/abs/2310.02704}}}
is inspired by that of Featherweight Generic
Go~\cite{featherweight-go}, but differs in some important aspects:
\begin{enumerate}
\item Methods are not included; instead we use \enquote{ordinary} top-level functions.
\item Go distinguishes syntactically between expressions and statements, whereas Featherweight Generic Go does not.
  We retain this distinction and discuss conversion between them in §\ref{sec:go:statements}.
\item Type parameters can be declared with an interface constraint. However, in
  our fragment the only available constraint is the unconstrained \go|any|, as
  Go's other constraints are not useful for our translation~(§\ref{sec:translation:dictionaries}).
\item We use modern Go's syntax for generics, which differs from the one used by
  Featherweight Generic Go, which pre-dates the introduction of generics in Go
  1.18 and was meant as a proposal demonstrating a possible design.
\end{enumerate}

\subsection{Declarations}

A (top-level) declaration $D$ can define either a new type or function.
Within one package, the order of declarations does not matter; any item may
reference any other. A program as a whole is simply a list $\overline{D}$ of
such declarations (note that we use overlines such as $\overline\alpha$ to
mean syntactic repetition).

\paragraph{Structure types}
A declaration of the form \go|type $t_S\big[\overline{\alpha\ c}\big]$ struct$\{\overline{A\ \tau}\}$| introduces a new type constructor with fields
$\overline{A}$ of types $\overline{\tau}$ to the program.
It may be polymorphic and take type arguments $\overline{\alpha}$ which can be freely referenced
within $\overline{\tau}$.
Since Go's syntax demands a constraint $c$ for each type variable $\alpha$, we always use \go|any|, which allows any type to be substituted for $\alpha$.

Note that there is no analogous construct to Thingol's sum types; that is, it is
not possible to a have a structure type which has more than one constructor.
Therefore, when encountering non-trivial sum types in Thingol, we must encode them accordingly (see~§\ref{sec:translation:data} for details).

\paragraph{Interface types}
A declaration of the form \go|type $t_I\big[\overline{\alpha\ c}\big]$ interface{}|
introduces a new (empty) interface type to the program.
While Go supports non-empty interfaces containing methods, we do not use
this feature (see §\ref{sec:translation:dictionaries}).

Unlike interfaces in typical object-oriented languages such as Java,
Go's interfaces are structural in nature:
any \go|struct| value conforms to an interface if (and only if)
the \go|struct| implements a superset of the declared methods of the interface.
This can also be probed at runtime.

This implies that empty interfaces correspond to a ``top'' type that can denote
arbitrary values. Go defines the unconstrained interface \go|any| as an alias
to this empty interface type,
which we use extensively in the translation scheme of data types, for reasons that will be explained later (§\ref{sec:translation:data}).
Additionally, we also use them for the translation of type classes (§\ref{sec:translation:dictionaries}).

\paragraph{Functions} A declaration \go|func $f\big[\overline{\alpha\ c}\big]$($\overline{x\; \tau}$) ($\overline{\gamma}$) { $s$ }| introduces
a new function $f$
to the program. The type parameters $\overline{\alpha}$ can be referenced within both argument
types $\overline{\tau}$ and the return types $\overline{\gamma}$.

Unlike in Thingol,
a function cannot have multiple equations nor perform pattern matching on its
arguments. Instead there is only one list of argument names $\overline{\alpha}$,
which are in scope for the (unique) function body $s$.

An unusual feature of Go is that its functions may return more than
one value (note that we have return types $\overline\gamma$ instead of just a
single return type $\gamma$):

\begin{lstlisting}[language=Go,basicstyle=\small\ttfamily]
func foo() (bool, int, string) {
  return false, 42, "bar"
}

func main() {
  x, y, z := foo()
}
\end{lstlisting}

\noindent At first glance this might seem analogous the tuples present in Standard ML, with
\go|foo()| returning a single value of the tuple \go|(bool, int, string)|. But this
is not the case; Go has no concept of tuples. Instead, the function itself returns
multiple values, which must be immediately assigned names (or discarded) at the
function's call site. Thus a call like \go|no_tuples := foo()| is not allowed.

\subsection{Expressions}

Expressions $e$ can have several forms: variables, function application, and function
abstraction are familiar from the $\lambda$-calculus. The others may require a bit more
explanation.

\paragraph{Structure literal} A literal of the form $t_s[\overline{\alpha}]\{\overline{e}\}$
gives a value of the \go|struct| type with name $t_S$ applied to type arguments
$\overline{\alpha}$, i.e., it produces a new value of the type $t_s\big[\overline{\alpha}\big]$
in which the fields are set to the evaluated forms of the expressions $\overline{e}$.
Note that the field names present in the
declaration of a \go|struct| type are absent: while they could be used, Go does
not require them. We omit them in the interest of shorter code.

\paragraph{Field selection} An expression $e.A$ selects the field named $A$ of
an expression $e$, which must have a fitting \go|struct| type $\tau_S$ that was
declared with a field name $A$, and returns the value of that field.
This is the only place outside a structure type's declaration that
field names are used.

\paragraph{Type conversion} An expression $\tau_I(e)$ evaluates to a value of
the interface type $\tau_I$ which contains the evaluated form of $e$ as its inner
value.
The original type $\sigma$ of $e$ is not erased at runtime;
it can be recovered using a type assertion statement (see the next section).
This expression can also be thought of as an ``upcast''.

\subsection{Statements}\label{sec:go:statements}

Unlike in Isabelle (and in Thingol) where \enquote{everything is an expression}, Go
has the same syntactic distinction between expression and statements that is common across
imperative languages: an assignment
\go|x := 42;| is a statement, not an expression, and cannot be used in places
where an expression is expected.

However, we constrain our fragment to only include sequences of statements that end in
a \go|return|.
This enables easy embedding of a statement into an expression: wrapping it into
an immediately-called lambda \go|func () $\tau$ { $s$ }()| suffices.
Note that a sequence of statements interspersed with \go|;|
is treated syntactically as a single statement.

The remainder of this section introduces the statement forms of our fragment.
All but the type assertion should feel familiar from similar languages.

\paragraph{Return} A statement \go|return $\overline{e}$| evaluates one or more
expressions, then returns from the current function. The $\overline{e}$ must match
the return types given in the function's head.

\paragraph{If statement} A statement of the form \go|if ($e$) { $s_1$ }; $s_2$|
will evaluate $e$, which must have a boolean type. If it
evaluates to the built-in value \go|true|, then $s_1$ is evaluated. Since
all statements end in \go|return|, it will then return from the current
function. Otherwise, $s_2$ is evaluated.
The form \go|if ($e$) { $s$ } else { $s_2$ }| would be semantically
equivalent within our fragment; we avoid it to reduce nesting in the generated
code.

\paragraph{Type assertion} A statement of the form \go|$x$, $y$ := $e$.($\sigma$)|
can be thought of as the inverse operation of type conversions, i.e., a ``downcast''.
For an expression $e$ of an interface type $\tau_I$, the assertion checks whether
the inner value contained within the interface value has type $\sigma$.
The boolean variable $y$ will indicate if the check was successful.
If so, $x$ will be bound to that inner value; otherwise, it will be \go|nil|, Go's null pointer.
Note that the type of $x$ is $\sigma$.

\section{Translation scheme}\label{sec:translation}

In this section, we will discuss the concrete translation schemes employed for
Thingol programs.
For brevity, we omit purely syntactic mappings, and focus on the non-trivial steps.

The translation scheme attempts to preserve names as far as possible.
Isabelle's Code Generator already provides (re)naming infrastructure, such as
generating guaranteed-unused \enquote{fresh} names where necessary.
In addition to that, some functions and data types require upper-case names, to match Go's
rules for exported symbols.

\subsection{Types, terms and statements}

We define three translations $\type(\tau)$, $\expr(t)$, and $\stmt(t)$.
The first is a straightforward syntactic mapping of types.
In the remainder of the chapter, we will informally equate Thingol types $\tau$
with their Go translation $\type(\tau)$ and write both simply as $\tau$.
For now, we exclude any mapping of common types (e.g.\ integers) to
built-in Go types; we will revisit this topic later (§\ref{sec:translation:mapping}).

The other two translations---$\expr$ and $\stmt$---are used for converting Thingol terms into Go expressions and statements.
Which one is used thus depends on what Go expects in each particular context;
for example, terms used as function arguments use $\expr$; a term which is a function body uses $\stmt$.
Semantically, for any term $t$, $\expr$ and $\stmt$ satisfy the following equivalences:
\begin{align*}
  \stmt(t) &\equiv \texttt{return}\; \expr(t) \texttt{;} \\
  \expr(t) &\equiv \texttt{func()}\; \tau \; \texttt{\{} \stmt(t) \texttt{\}()}
\end{align*}

\paragraph{Abstractions}

The translation of a $\lambda$-abstraction
$\lambda (x :: \tau).\; (t :: \gamma)$ demonstrates the distinction
between expressions and statements:
\[\expr(\lambda (x :: \tau).\; (t :: \gamma))
    = \texttt{func (} x\;\tau\texttt{)}\ \gamma\ \texttt{\{} \stmt(t) \texttt{\}}\]
Although curried abstractions are unusual in Go, no effort is made to uncurry
them (unlike top-level functions, which we do uncurry~§\ref{sec:translation:top-level-functions}).

\paragraph{Applications of top-level functions}

Applications $t$ are more tedious:
Definitions of top-level functions are uncurried (§\ref{sec:translation:top-level-functions}),
so we first have to check if $t$ is a call to such a function, i.e., if
$t$ has the shape $\big(\cdots\left((f[\overline{\tau}_i]\; a_1)\; a_2\right)\cdots\big)\; a_n$,
where $f$ references a top-level function or data type constructor taking $m$ arguments.

If so, we have to consider three cases:
\begin{enumerate}
  \item Fully-satured application ($n = m$); all arguments are passed into $f$
  \item Unsatured application ($n < m$); need to $\eta$-expand
  \item Over-satured application ($n > m$). This occurs if $f$ returns another function,
    with $a_1$ to $a_m$ being the immediate arguments to $f$ and any remaining
    $a_{m+1}$ to $a_n$ as curried arguments. The latter will be passed individually.
\end{enumerate}
As will be described later~(§\ref{sec:translation:dictionaries}),
the dictionary construction used to encode Isabelle's type classes
may introduce additional (value-level) parameters to top-level functions,
also adding corresponding additional arguments $d_1$ to $d_r$ to each of their
applications. These are inserted before the user-defined parameters.

Altogether, we arrive at the following scheme when $f$ references a function:
\[\expr(t)
   = \mbox{\go|$f$[$\tau_1$,$\dots$,$\tau_i$]($d_1$,...$d_r$,$a_1$,...,$a_m$)($a_{m+1}$)...($a_n$)|}\]
Finally, if $f$ references a data type constructor of a type $\tau$ rather than a function,
the case $n > m$ cannot occur.
However, we must wrap the constructor into a type conversion to type $\tau$,
and use slightly different syntax for passing the arguments:
\[\expr(t)
   = \mbox{\go|$\kappa$($f$[$\tau_1$,$\dots$,$\tau_i$]\{$d_1$,...$d_r$,$a_1$,...,$a_m$\})|}\]

\paragraph{Lambda applications}

If an application $t = t_1 \; t_2$ is not a call to a top-level function, then the
translation is straightforward:
$\expr(t_1\; t_2) = \expr(t_1)\texttt{(}\expr(t_2)\texttt{)}$.

\subsection{Data types}\label{sec:translation:data}

A data type $\kappa$ defined in Thingol consists of type
parameters $\overline{\alpha}_i$ and constructors $\overline{f}_i$.
Each $f_i$ gets translated into its own separate \go|struct| type.

As was discussed in §\ref{sec:go}, Go knows no sum types, thus the translation
has to simulate their behaviour by other means. For a data type, we generate
a new unconstrained interface type $\delta$, meant to represent any
constructor $f_i$ of $\kappa$.

If the data type $\kappa$ has exactly one constructor $f_1$, then no additional
interface type $\delta$ is generated.

\paragraph{Constructors}

Defining a \go|struct| type for an individual constructor is straightforward.
A constructor $f$ with fields of types $\tau_1$ to $\tau_i$ is translated
into Go as a \go|struct| with the same name and fields:
\go|type $f$ struct {$\overline{A\; \tau}_i$}|,
where the $\overline{A}_i$ are newly-invented names for each of the fields, as
no field names are present in Thingol.
Note that those generated field names are entirely unimportant (access happens
only through destructors, and the names are not required when constructing a
value); the only requirement imposed on them is that each $\overline{A}_i$ of
the same \go|struct| are distinct. Thus the type \lstinline|Nat| (Figure~\ref{listing:thingol})
becomes:
\begin{lstlisting}[basicstyle=\ttfamily\small]
type Nat any;
type Zero struct { };
type Suc struct { A Nat; };
\end{lstlisting}

\noindent With that, we can construct the number 1 as \go|Nat(Suc{Nat(Zero{}))|.
The interface type $\delta$ (here \go|Nat|) acts as a faux sum type: the translation promises
that (as long as its input program was type-correct) it will never contain
anything but values of types \go|Zero| and \go|Suc|. On the Go side, there is no
such guarantee: it sees \go|Nat| as unconstrained, and would happily allow such
values as \go|Nat(Suc{nil})| or even \go|Suc{"wrong"}|, leading to runtime
exceptions elsewhere in the generated code, especially in translated pattern
matches~(§\ref{sec:translation:cases}).

\paragraph{Destructors}

Along with each constructor's \go|struct| type, we generate a synthetic
function $\dest{f}$ not present in the Thingol program, to be used as a destructor
in the translation of Thingol's case expressions (§\ref{sec:translation:cases}).
Their sole purpose is to unpack and return the individual fields in a \go|struct| type,
exploiting Go's multiple return types.

\begin{lstlisting}[basicstyle=\ttfamily\small]
func $\dest{f}$ (p $f$) ($\tau_1$, $\dots$, $\tau_n$) {
  return p.$A_1$, $\dots$, p.$A_n$
}
\end{lstlisting}
Destructors are omitted when there are no fields to unpack. For \lstinline|Nat|,
we need only one:
\begin{lstlisting}[basicstyle=\ttfamily\small]
func Suc_dest(p Suc)(Nat) { return p.A; }
\end{lstlisting}

\paragraph{Example} Slightly more involved is the \lstinline|$\alpha$ list|
data type (Figure~\ref{listing:thingol}). It is polymorphic, and thus requires use of Go's generics:
\begin{lstlisting}[basicstyle=\ttfamily\small]
type List[a any] interface {};
type Nil[a any] struct { };
type Cons[a any] struct { A a; Aa List[a]; };
func Cons_dest[a any](p Cons[a])(a, List[a]) { return p.A, p.Aa; }
\end{lstlisting}

\subsection{Case expressions}\label{sec:translation:cases}

Thingol's case expressions implement pattern matching on a value, in a way
which will be immediately familiar from other functional
languages such as Standard ML or Haskell: they inspect a \emph{scrutinee} $t$
and match it against a series of clauses
$\overline{p_i \rightarrow b_i}$.
Each clause contains a pattern $p_i$ and a term $t_i$ that is to be evaluated
if the pattern matches the scrutinee.
Syntactically, patterns are a subset of terms; they can only be composed of
variables and fully-satisfied applications of data type constructors to
sub-patterns $f \; \overline{p}_i$ constructed of the same subset.

Since Go has no comparable feature, a data type pattern in a case expression is
translated into a series of (possibly nested) \go|if|-conditions and calls
to destructor functions. The bodies of the innermost \go|if|-condition then correspond to the translated terms $t_i$, which must be in statement-form, i.e., ending in
a \go|return|-statement.
Thus, if the pattern could be matched, further patterns
will not be executed.
Naturally, using \go|return| in this manner implies that a case expression must always either be in tail position, or
else be wrapped into an anonymous function if it does not (§\ref{sec:go}).

If the pattern did not match, execution
will continue with either the next block of \go|if|-conditions generated from
the next clause, or encounter a final catch-all call to Go's
built-in \go|panic| function, which aborts the program in case of an incomplete
pattern where no clause could be matched (incomplete patterns are admissable
in Isabelle's logic, see Hupel~\cite{hupel-dicts} for a detailed description).
This \go|panic| can also be encountered if an external caller exploited the lossy conversion
of sum types as described above and supplied, e.g., a \go|nil| value as a scrutinee.

Taken together, an entire case expression is translated as a linear sequence of individual clauses, followed by a \go|panic|:
\[
  \stmt(\text{case } t :: \tau \text{ of }  [\overline{p \rightarrow b}]) = \overline{\stmt(p \rightarrow b); } \; \mbox{\go|panic("Match failed"); |}
\]
Let us now consider the concrete translation for variable and constructor patterns.

\paragraph{Variable pattern}

We assign the scrutinee $t$ to the variable $x$ to make it available in the scope of $b$:
$\stmt(x \rightarrow b) = \texttt{\{} x\; \texttt{:=}\;\expr(t)\texttt{;}\; \stmt(b)\texttt{\}}$.

\paragraph{Constructor pattern}

The pattern is of the form $f [\overline{\tau}_i] [\overline{s}_k]$.
If all sub-patterns $\overline{s}_k$ are variable patterns, the translation is once again straightforward:%
\begin{align*}
  \stmt(f [\overline{\tau}_i] [\overline{s}_k] \rightarrow b) =
    \texttt{\{}& q, m \texttt{:=}\; t.\texttt{(}f[\overline{\tau}_i]\texttt{);}\;
    \\
    & \texttt{if (}m\texttt{) \{} A_1, \dots, A_k \;\texttt{:=}\; \dest{f}(t)\texttt{;}\;
      \stmt(b) \texttt{\}\}}
\end{align*}
Nested constructor patterns are translated in the same way, but pushed inwards into the
body of the \go|if|-statement generated above:
\begin{align*}
  \stmt(f [\overline{\tau}_i] [\overline{s}_k] \rightarrow b) &=
    \texttt{\{} q, m \texttt{:=}\; t.\texttt{(}f[\overline{\tau}_i]\texttt{);}\;
    \\
    & \qquad\texttt{if (}m\texttt{) \{} A_1, \dots, A_k \;\texttt{:=}\; \dest{f}(t)\texttt{;}\;
     \;\mathcal I\; \texttt{\}\}}
  \\
  \mathcal I &= \stmt(\text{case } A_1 \text{ of } s_1 \rightarrow ( \ldots \rightarrow (\text{case } A_k \text{ of } s_k \rightarrow b)))
\end{align*}
In other words, the sub-patterns are treated as if they were further nested case expressions.
This results in a total nesting depth of one level per constructor.

Within the innermost \go|if|, the body $b$ of the pattern's clause is translated as
statement to ensure it returns from the current function.

\paragraph{Optimizing the nesting level}
The translation described in this section can translate arbitrary patterns, but comes at the price
of potentially exponential code blow-up. Even a single pattern consisting of
just a constructor and $k$ fields, none of which are proper patterns, will still
produce $k$ levels of nested \go|if|-statements. But if the fields themselves
are again data type constructors with sub-patterns, the number of nested levels
quickly increases further.

In real-world applications, we can reduce the blow-up by optimizing
constructor patterns without arguments.
Instead of calling a destructor function, we can emit an equality
check, since there are no fields to extract.
Multiple equality checks can be joined together using Go's conjunction operator \go|&&|.

\paragraph{Example}
Consider the function \texttt{hd2} (Figure~\ref{listing:thingol}), which takes a list and returns (optionally) the second element of the list. It is translated into Go as follows:
\begin{lstlisting}[basicstyle=\ttfamily\small]
func Hd2[a any] (x0 List[a]) Option[a] {
  if (x0 == (List[a](Nil[a]{}))) {
    return (Option[a](None[a]{}));
  }
  q, m := x0.(Cons[a]);
  if (m) {
    _, c := Cons_dest(q);
    if (c == (List[a](Nil[a]{}))) {
      return (Option[a](None[a]{}));
    }
  }
  q, m := x0.(Cons[a]);
  if (m) {
    _, p := Cons_dest(q);
    q, m := p.(Cons[a]);
    if (m) {
      ya, _ := Cons_dest(q);
      return (Option[a](Some[a]{ya}));
    }
  }
  panic("match failed");
}
\end{lstlisting}
This piece of generated code benefits from the optimisation described above (in its first two clauses).
Also, observe that since unused variables are a compile error in Go, unused bound
names above have been generated as \go|_| instead.

\subsection{Top-level functions}\label{sec:translation:top-level-functions}

Unlike lambdas that occur within terms, top-level functions in Thingol can
have multiple clauses and pattern-match on their arguments, neither of which is
supported in Go. It is thus necessary to translate them differently: all
equations of the same function will have to be merged, with the pattern matching
on their parameters again pushed inwards into the then combined, single
function body.

Further, treating them differently from in-term lambda expression also allows the
generator to uncurry them, creating code that is much closer to an idiomatic
style in Go.

\paragraph{Merging multiple clauses}
Thingol allows Haskell-style function definition comprising multiple clauses.
But in Go, all parameters of functions must be simple variables.
Thus, if any of the parameters patterns $\overline{p_i}$ is a
proper pattern, a fresh name $x_i$ for it is invented. Likewise, if a parameter is
a variable binding instead of a proper pattern, but has multiple different
names in two clauses, the name $x_i$ used in the first clause is picked as the name
of the parameter in Go.

\paragraph{Pattern matching}
The combined function body then contains a pattern match translation, as described above.%
\footnote{The
already-existing Scala target uses a similar transformation.}
Each equation is
treated as a clause of a synthetic case-expression; for functions matching
on multiple parameters, we again push inwards and
translate as if a nested series of case-expressions were present.

\paragraph{Example}
Consider this definition for \lstinline|hd2'|, which is semantically equivalent
to \lstinline|hd2|, but written using multiple equations:
\begin{lstlisting}[basicstyle=\ttfamily\small]
fun hd2' :: $\forall \alpha. \alpha\;\mathtt{list} \Rightarrow \alpha\;\mathtt{option}$ where
  hd2' Nil = None
  hd2' (Cons $x$ Nil) = None
  hd2' (Cons $x$ (Cons $y$ $\mathit{xs}$) $\mathit{ys}$) = Some $y$
\end{lstlisting}
The generated Go code is identical.

\paragraph{Special case: top-level constants}
Thingol accepts top-level definitions that are not functions, for example:
\lstinline|definition a :: nat where a = 10|.
Unfortunately, Go admits top-level variable declarations only for monomophic types,
and further disallows function calls in their definitions.

Therefore, we must treat such Thingol definitions as if they were nullary functions.
While this changes nothing of the semantics of the translated program, it does
incur a (potentially significant) runtime cost:
constants will be evaluated each time they are used, instead of only once when the program
is initialized.

\subsection{Dictionary construction}\label{sec:translation:dictionaries}

On the surface, Isabelle's Haskell-style type classes and Go's interfaces share many of the same features, and are sometimes
considered to be near-analogous~\cite{generic-go-to-go}.
However, translating type classes into interfaces does not work, due to an
implementation concern: Go directly compiles methods into virtual tables for dynamic dispatch.
A Go interface declares multiple \emph{methods}, where each method type must take the generic value as its zeroth (i.e.\ implicit) parameter.
Thingol has no such restriction. Consider, for example:

\noindent
\begin{minipage}{.5\linewidth}
\begin{lstlisting}[basicstyle=\small\ttfamily]
class foo where
  foo :: unit $\Rightarrow$ $\alpha$
\end{lstlisting}
\end{minipage}
\begin{minipage}{.5\linewidth}
\begin{lstlisting}[basicstyle=\small\ttfamily]
class bar where
  bar :: ($\alpha$ $\Rightarrow$ $\alpha$) $\Rightarrow$ unit
\end{lstlisting}
\end{minipage}

\noindent As Go interfaces, both are invalid:
\lstinline|foo| declares a function whose parameter types do not mention $\alpha$ at all, while \lstinline|bar|'s function does not take a simple $\alpha$ parameter (but a parameter whose type contains $\alpha$).

To avoid the additional complexity of treating all these cases separately,
we resort to using a dictionary construction~\cite{hupel-dicts,haftmann-nipkow}
in all cases. Since the existing SML target of the Code Generator has to deal
with the same issue, the required infrastructure is already in place:
Thingol's terms come with enough annotations to resolve all type class constraints during
translation and replace the implicit instance arguments of functions making use
of type classes by explicit dictionary values, which we represent as one data
type per type class.

Thus only relatively few things are left to do in Go:
\begin{enumerate}
\item declare a data type for each type class, called its \emph{dictionary} type
\item translate type class constraints on functions into explicit function
  arguments of dictionary types
\item translate type class instances into either a value of the type class's
  dictionary type, or, if the instance itself takes type class constraints, to a
  function producing such a value when given values of dictionary types representing
  these constraints
\item any time a top-level function is used, the already-resolved type class
  constraints must be given as explicit arguments
\end{enumerate}

\paragraph{Example} The class declarations (Figure~\ref{listing:thingol}) are translated as follows:
\begin{lstlisting}[basicstyle=\small\ttfamily]
type Semigroup[a any] struct {
  Plus func(a, a) a
}

type Monoid[a any] struct {
  Semigroup_monoid Semigroup[a]
  Zero func () a
}

func Sum[a any] (a_ Monoid[a], xs List[a]) a {
  return Fold[a, a](
    func (aa a) func(a) a {
      return func (b a) a { return a_.Semigroup_monoid.Plus(aa, b); };
    },
    xs, a_.Zero()
  );
}
\end{lstlisting}
\subsection{Mapping high-level constructs}\label{sec:translation:mapping}

So far, the shallow embedding we have presented produced code with no dependencies
on the Go side, with only the built-in constructs \go|panic| and \go|&&| used.
All higher-level constructs used by programs (such as lists, numbers) must thus
be \enquote{brought along} from Isabelle, and are translated wholesale exactly
as they are defined in their formalisations. While this guarantees correctness,
it is highly impractical for real-world applications: for example, natural numbers
as defined in Isabelle/HOL (unary Peano representation, §\ref{sec:translation:data}) require linear memory
and quadratic runtime even for simple operations like addition.

Luckily, the Code Generator already has a solution for this conundrum in the
form of \emph{printing rules}, which can map Isabelle's types and constants to
user-supplied names in the target language. We have set up printing rules mapping:
\begin{itemize}
  \item Isabelle/HOL's booleans to booleans in Go
  \item numbers to arbitrary-precision integers (via Go's \go|math/big| package)
  \item strings of the \texttt{String.literal} type to strings in Go
\end{itemize}
Unfortunately, linked lists cannot be as easily mapped by default, because Go does not feature a
standard implementation of linked lists.

\section{Evaluation}\label{sec:evaluation}

Even though Go greatly differs from the existing targets, we have achieved almost
full feature parity with the translation described in this paper. Isabelle
constructs that are not (cleanly) mapped are:

\begin{itemize}
  \item infinite data types, which can be defined e.g.\ via \lstinline|codatatype| in Isabelle, but are rejected by Go's compiler;
  \item some low-level string operations that operate on byte values of characters.
\end{itemize}

\paragraph{Trusted code base}
All target language generators are part of Isabelle's \emph{trusted code base},
i.e.\ bugs inside its own code may lead to bugs in the generated program, and
cannot be checked for by Isabelle's kernel.
Fortunately, our implementation is ``just another module'' to the core
infrastructure; up until Thingol everything remains unchanged, in line with the other language targets.

However, future (more ambitious) code printing may require changes in Thingol:
If code printing shall assume more constructs of Go, it would be useful for
Thingol itself to have some concept of the syntactic distinction between expressions and statements.

\paragraph{Code style}
The generated Go code is not idiomatic, but neither is the generated code for the other languages.
Even though the semantics of SML, OCaml and others may more closely resemble the intention of Isabelle users, the generated code in those languages is also littered with syntactic artifacts.
This is evidenced by the fact that neither SML nor OCaml support type classes, and Scala code hardly uses type classes in the way that Haskell does (typically prefering object-orientation).
Therefore, we do not envision a future need to align the style of the generated code more closely with the preferred style of hand-written Go code.

The main challenge arises from interfacing between generated and hand-written
Go code, both of which would be present in a typical application. For instance,
constructing
values for the translated \lstinline|datatype| definitions or using curried
functions in Go is unfortunately verbose, and can easily introduce errors.

We therefore recommend to write wrapper code that exposes a \enquote{cleaner}
interface, ready to be consumed by the real-world application.
The wrapper must be written carefully:
many explicit type annotations are needed in the code, and not all incorrect type annotations will cause
compilation to fail. In particular, if a data type's constructor
is annotated with a wrong \go|interface| type, the assumption underlying the
translation of case-expressions will fail, resulting in a \enquote{match failed}
error at runtime~(§\ref{sec:translation:cases}).

Another awkward source of problems when integrating the generated code with
a larger code base is that Go's standard library lacks common functional data
structures, such as lists or tuples~(§\ref{sec:translation:mapping}).
Hand-written code would need to deal with the necessary conversions
(e.g. from a Go array into a linked list). To some extent, this can be
alleviated by leveraging third-party libraries for functional data structures,
which are unfortunately not popular in the Go community.

\subsection{Case studies}

We conducted two case studies that have confirmed our approach.

\paragraph{Existing formalisation}

At Giesecke+Devrient, we use Isabelle for a substantial formalisation of various graph algorithms
powering a financial transaction system. The purpose of the formalisation is to
provide real-world security guarantees, such as inability to clone money.
We have previously used the Code Generator to produce Scala code as a reference
implementation, combined with some hand-written wrapper code and basic unit tests.

As a simple evaluation of Go code generated from the same Isabelle theories,
we re-wrote the unit tests and the necessary wrapper code in Go.
We obtained equivalent results and could not find bugs in the Code Generator or unintended
behaviour of the code it produced.
Note that no adaptations of the Isabelle formalisation were necessary,
which proves that the Go backend works as a drop-in replacement for the other targets.

Starting from this, we can narrow the formalisation gap mentioned in the
introduction. It allows us to link the Isabelle/HOL reference implementation
with the real-world production implementation in Go.

\paragraph{HOL-Codegenerator$\_$Test}

Isabelle's distribution contains a Code Generator test session which
is used as a self-check for the various target languages of the Code Generator.
For this paper, a single export command is relevant, which is meant to export
a considerable chunk of Isabelle/HOL's library as a stress-test for the Code Generator.
This has worked as expected, with the entirety of the test suite successfully
compiling in Go.

As a consequence, our approach enables the vast majority of Isabelle users to generate Go code without having to rewrite their formalisation.
In particular---because we map to a functional fragment of Go---there is no need for users to reach for a deep embedding of an imperative language.

\section{Conclusion}

We have presented a translation from Thingol by shallow embedding into a fragment
of Go, and implemented it as a target language for Isabelle's code generation framework.
The new target language has been used with success to port an existing Isabelle
formalisation that was only targeting Scala to additionally target Go.
The implementation is readily usable with a standard Isabelle2024 installation and
requires merely importing an additional theory file.
The suite of existing tests of Isabelle's Code Generator is also supported.

\paragraph{Future work}
The two most promising areas of future work are: leveraging Go's imperative nature
by tightly integrating it with Imperative/HOL~\cite{imperative-hol}; and generating
code that utilizes more of Go's standard libraries through custom code printing rules.
Both can be implemented using similar mechanisms. However, substantial changes to Isabelle's code
generation infrastructure are required, because Go demands more type annotations
than other target languages.

\subsubsection{Acknowledgements}
The authors would like to thank Florian Haftmann and Cornelius Diekmann for their contributions to the development.
This work has been partially supported by the Federal Ministry of Education and Research (BMBF), Verbundprojekt CONTAIN (13N16582).

\subsubsection{Availability}
The artifact for this paper is available in the \emph{Archive of Formal Proofs (AFP)} \cite{Go-AFP}
and under the DOI \url{https://doi.org/10.5281/zenodo.11608252}.

\bibliographystyle{splncs04}
\bibliography{references}

\appendix

\section{Syntax of functional Go}\label{sec:go-syntax}

Overlines (such as $\overline\alpha$) denote syntactic repetition.

\begin{longtable}[l]{ll}
Field name & $A$ \\
Function name & $f$ \\
Variable name & $x$, $y$ \\
Structure type name & $t_S$ \\
Interface type name & $t_I$ \\
Type parameter & $\alpha$ \\
Type name & $t ::= t_S\tinyskip |\tinyskip t_I$ \\
Type & $\tau, \gamma, \sigma ::= $ \\
\tinyskip Parameter & \tinyskip $\alpha$ \\
\tinyskip Structure Type & \tinyskip $t_S\big[\overline{\tau}\big]$ \\
\tinyskip Interface Type & \tinyskip $t_I\big[\overline{\tau}\big]$ \\
Function head & $F ::= $\go|($\overline{x\tinyskip \tau}$)| $(\overline{\tau})$ \\
Type literal & $T ::=$ \\
\tinyskip Structure & \tinyskip \go|struct{$\overline{A\tinyskip \tau}$}| \\
\tinyskip Interface & \tinyskip \go|interface{}| \\
Interface constraint & $c ::=\ $\go|any|\\
Declaration & $D ::=$ \\
\tinyskip Type declaration & \tinyskip\go|type $t[\overline{\alpha\ c}]$ $T$| \\
\tinyskip Function declaration & \tinyskip\go|func $f[\overline{\alpha\ c}]\ F$ { $s$ }| \\
Expression & $d, e ::=$ \\
\tinyskip Variable & $\tinyskip x$ \\
\tinyskip Function call & $\tinyskip f[\overline{\alpha}](\overline{e})$ \\
\tinyskip Structure literal & $\tinyskip t_s[\overline{\alpha}]\{\overline{e}\}$ \\
\tinyskip Function abstraction & $\tinyskip$\go|func $F$ { $s$ }| \\
\tinyskip Field selection & $\tinyskip e.A$ \\
\tinyskip Type conversion & $\tinyskip \tau_I(e)$ \\
\tinyskip Null pointer & \tinyskip\go|nil| \\
Statement & $s ::=\ $ \\
\tinyskip Expression & \tinyskip\go|return $\overline{e}$;| \\
\tinyskip Variable declaration & \tinyskip\go|$x$ := $e$; $s$| \\
\tinyskip if-Statement & \tinyskip\go|if ($e$) { $s$ }; $s$| \\
\tinyskip Type assertion & \tinyskip\go|$x$, $y$ := $e$.($\tau$); $s$|
\end{longtable}

\section{Example: Red-black trees}\label{sec:rbtree}

\subsection{Isabelle specification}

First, let us consider a small subset of the definition of red-black trees in
Isabelle/HOL. Isabelle's syntax admits abbreviations that are expanded both on
the left-hand and right-hand sides of defining equations, leading to a very
compact definition of balancing.

\begin{lstlisting}
datatype 'a tree =
  Leaf |
  Node ('a tree) 'a ('a tree)

datatype color = Red | Black

type_synonym 'a rbt = ('a * color) tree

abbreviation R where R l a r == Node l (a, Red) r
abbreviation B where B l a r == Node l (a, Black) r

fun baliL :: 'a rbt => 'a => 'a rbt => 'a rbt where
baliL (R (R t1 a t2) b t3) c t4 = R (B t1 a t2) b (B t3 c t4) |
baliL (R t1 a (R t2 b t3)) c t4 = R (B t1 a t2) b (B t3 c t4) |
baliL t1 a t2 = B t1 a t2
\end{lstlisting}

\subsection{Scala translation (for reference)}

As a point of reference, consider the generated Scala code. Note that the data type
definitions in Isabelle/HOL are rather compact. But hidden behind the scenes, HOL's implementation
of recursive functions will always expand pattern matching to disambiguate
patterns. The advantage is that patterns are no longer sensitive about the
order in which they are applied. The disadvantage however is that the number of
patterns may explode, depending on their depth. This is well-documented
behaviour and is independent of code generation~\cite{krauss-fun}.

\begin{lstlisting}[basicstyle=\scriptsize\ttfamily]
object RbtTest {

abstract sealed class color
final case class Red() extends color
final case class Black() extends color

abstract sealed class tree[A]
final case class Leaf[A]() extends tree[A]
final case class Node[A](a: tree[A], b: A, c: tree[A]) extends tree[A]

def baliL[A](x0: tree[(A, color)], c: A, t4: tree[(A, color)]): tree[(A, color)]
  =
  (x0, c, t4) match {
  case (Node(Node(t1, (a, Red()), t2), (b, Red()), t3), c, t4) =>
    Node[(A, color)](Node[(A, color)](t1, (a, Black()), t2), (b, Red()),
                      Node[(A, color)](t3, (c, Black()), t4))
  case (Node(Leaf(), (a, Red()), Node(t2, (b, Red()), t3)), c, t4) =>
    Node[(A, color)](Node[(A, color)](Leaf[(A, color)](), (a, Black()), t2),
                      (b, Red()), Node[(A, color)](t3, (c, Black()), t4))
  case (Node(Node(v, (vc, Black()), vb), (a, Red()), Node(t2, (b, Red()), t3)),
         c, t4)
    => Node[(A, color)](Node[(A, color)](Node[(A, color)](v, (vc, Black()), vb),
  (a, Black()), t2),
                         (b, Red()), Node[(A, color)](t3, (c, Black()), t4))
  case (Leaf(), a, t2) => Node[(A, color)](Leaf[(A, color)](), (a, Black()), t2)
  case (Node(Leaf(), (v, Black()), vb), a, t2) =>
    Node[(A, color)](Node[(A, color)](Leaf[(A, color)](), (v, Black()), vb),
                      (a, Black()), t2)
  case (Node(Leaf(), va, Leaf()), a, t2) =>
    Node[(A, color)](Node[(A, color)](Leaf[(A, color)](), va,
                                       Leaf[(A, color)]()),
                      (a, Black()), t2)
  case (Node(Leaf(), va, Node(v, (ve, Black()), vd)), a, t2) =>
    Node[(A, color)](Node[(A, color)](Leaf[(A, color)](), va,
                                       Node[(A, color)](v, (ve, Black()), vd)),
                      (a, Black()), t2)
  case (Node(Node(vc, (vf, Black()), ve), (v, Black()), vb), a, t2) =>
    Node[(A, color)](Node[(A, color)](Node[(A, color)](vc, (vf, Black()), ve),
                                       (v, Black()), vb),
                      (a, Black()), t2)
  case (Node(Node(vc, (vf, Black()), ve), va, Leaf()), a, t2) =>
    Node[(A, color)](Node[(A, color)](Node[(A, color)](vc, (vf, Black()), ve),
                                       va, Leaf[(A, color)]()),
                      (a, Black()), t2)
  case (Node(Node(vc, (vf, Black()), ve), va, Node(v, (vh, Black()), vg)), a,
         t2)
    => Node[(A, color)](Node[(A, color)](Node[(A,
        color)](vc, (vf, Black()), ve),
  va, Node[(A, color)](v, (vh, Black()), vg)),
                         (a, Black()), t2)
  case (Node(v, (vc, Black()), vb), a, t2) =>
    Node[(A, color)](Node[(A, color)](v, (vc, Black()), vb), (a, Black()), t2)
}

}
\end{lstlisting}

\subsection{Go translation}

The generated Go code has roughly four times the size as the reference Scala
code. This is explained by the constant overhead in the data type definitions,
as well as the overhead for encoding pattern matching.

\begin{lstlisting}[language=Go,basicstyle=\scriptsize\ttfamily,breaklines=true]
package RbtTest

import (
)

// sum type which can be Red, Black
type Color any;
type Red struct { };
type Black struct { };



// sum type which can be Leaf, Node
type Tree[a any] any;
type Leaf[a any] struct { };
type Node[a any] struct { A Tree[a]; Aa a; Ab Tree[a]; };

func Node_dest[a any](p Node[a])(Tree[a], a, Tree[a]) {
  return p.A, p.Aa, p.Ab
}

type Prod[a, b any] struct { A a; Aa b; };
func Pair_dest[a, b any](p Prod[a, b])(a, b) {
  return  p.A, p.Aa;
}

func BaliL[a any] (x0 Tree[Prod[a, Color]], c a, t4 Tree[Prod[a, Color]]) Tree[Prod[a, Color]] {
  {
    q, m := x0.(Node[Prod[a, Color]]);
    if m {
      q, p, t3a := Node_dest(q);
      r, m := q.(Node[Prod[a, Color]]);
      if m {
        t1a, r, t2a := Node_dest(r);
        _ = r;
        ab, d := Pair_dest(r);
        if d == (Color(Red{})) {
          _ = p;
          ba, e := Pair_dest(p);
          if e == (Color(Red{})) {
            cb := c;
            t4b := t4;
            return Tree[Prod[a, Color]](Node[Prod[a, Color]]{Tree[Prod[a, Color]](Node[Prod[a, Color]]{t1a, Prod[a, Color]{ab, Color(Black{})}, t2a}), Prod[a, Color]{ba, Color(Red{})}, Tree[Prod[a, Color]](Node[Prod[a, Color]]{t3a, Prod[a, Color]{cb, Color(Black{})}, t4b})});
          }
        }
      }
    }
  };
  {
    q, m := x0.(Node[Prod[a, Color]]);
    if m {
      d, q, p := Node_dest(q);
      if d == (Tree[Prod[a, Color]](Leaf[Prod[a, Color]]{})) {
        _ = q;
        ab, e := Pair_dest(q);
        if e == (Color(Red{})) {
          r, m := p.(Node[Prod[a, Color]]);
          if m {
            t2a, r, t3a := Node_dest(r);
            _ = r;
            ba, f := Pair_dest(r);
            if f == (Color(Red{})) {
              cb := c;
              t4b := t4;
              return Tree[Prod[a, Color]](Node[Prod[a, Color]]{Tree[Prod[a, Color]](Node[Prod[a, Color]]{Tree[Prod[a, Color]](Leaf[Prod[a, Color]]{}), Prod[a, Color]{ab, Color(Black{})}, t2a}), Prod[a, Color]{ba, Color(Red{})}, Tree[Prod[a, Color]](Node[Prod[a, Color]]{t3a, Prod[a, Color]{cb, Color(Black{})}, t4b})});
            }
          }
        }
      }
    }
  };
  {
    q, m := x0.(Node[Prod[a, Color]]);
    if m {
      r, q, p := Node_dest(q);
      s, m := r.(Node[Prod[a, Color]]);
      if m {
        va, s, vba := Node_dest(s);
        _ = s;
        vca, d := Pair_dest(s);
        if d == (Color(Black{})) {
          _ = q;
          ab, e := Pair_dest(q);
          if e == (Color(Red{})) {
            t, m := p.(Node[Prod[a, Color]]);
            if m {
              t2a, t, t3a := Node_dest(t);
              _ = t;
              ba, f := Pair_dest(t);
              if f == (Color(Red{})) {
                cb := c;
                t4b := t4;
                return Tree[Prod[a, Color]](Node[Prod[a, Color]]{Tree[Prod[a, Color]](Node[Prod[a, Color]]{Tree[Prod[a, Color]](Node[Prod[a, Color]]{va, Prod[a, Color]{vca, Color(Black{})}, vba}), Prod[a, Color]{ab, Color(Black{})}, t2a}), Prod[a, Color]{ba, Color(Red{})}, Tree[Prod[a, Color]](Node[Prod[a, Color]]{t3a, Prod[a, Color]{cb, Color(Black{})}, t4b})});
              }
            }
          }
        }
      }
    }
  };
  {
    if x0 == (Tree[Prod[a, Color]](Leaf[Prod[a, Color]]{})) {
      ab := c;
      t2a := t4;
      return Tree[Prod[a, Color]](Node[Prod[a, Color]]{Tree[Prod[a, Color]](Leaf[Prod[a, Color]]{}), Prod[a, Color]{ab, Color(Black{})}, t2a});
    }
  };
  {
    q, m := x0.(Node[Prod[a, Color]]);
    if m {
      d, p, vba := Node_dest(q);
      if d == (Tree[Prod[a, Color]](Leaf[Prod[a, Color]]{})) {
        _ = p;
        va, e := Pair_dest(p);
        if e == (Color(Black{})) {
          ab := c;
          t2a := t4;
          return Tree[Prod[a, Color]](Node[Prod[a, Color]]{Tree[Prod[a, Color]](Node[Prod[a, Color]]{Tree[Prod[a, Color]](Leaf[Prod[a, Color]]{}), Prod[a, Color]{va, Color(Black{})}, vba}), Prod[a, Color]{ab, Color(Black{})}, t2a});
        }
      }
    }
  };
  {
    q, m := x0.(Node[Prod[a, Color]]);
    if m {
      e, vaa, d := Node_dest(q);
      if e == (Tree[Prod[a, Color]](Leaf[Prod[a, Color]]{})) && d == (Tree[Prod[a, Color]](Leaf[Prod[a, Color]]{})) {
        ab := c;
        t2a := t4;
        return Tree[Prod[a, Color]](Node[Prod[a, Color]]{Tree[Prod[a, Color]](Node[Prod[a, Color]]{Tree[Prod[a, Color]](Leaf[Prod[a, Color]]{}), vaa, Tree[Prod[a, Color]](Leaf[Prod[a, Color]]{})}), Prod[a, Color]{ab, Color(Black{})}, t2a});
      }
    }
  };
  {
    q, m := x0.(Node[Prod[a, Color]]);
    if m {
      d, vaa, p := Node_dest(q);
      if d == (Tree[Prod[a, Color]](Leaf[Prod[a, Color]]{})) {
        q, m := p.(Node[Prod[a, Color]]);
        if m {
          vb, q, vda := Node_dest(q);
          _ = q;
          vea, e := Pair_dest(q);
          if e == (Color(Black{})) {
            ab := c;
            t2a := t4;
            return Tree[Prod[a, Color]](Node[Prod[a, Color]]{Tree[Prod[a, Color]](Node[Prod[a, Color]]{Tree[Prod[a, Color]](Leaf[Prod[a, Color]]{}), vaa, Tree[Prod[a, Color]](Node[Prod[a, Color]]{vb, Prod[a, Color]{vea, Color(Black{})}, vda})}), Prod[a, Color]{ab, Color(Black{})}, t2a});
          }
        }
      }
    }
  };
  {
    q, m := x0.(Node[Prod[a, Color]]);
    if m {
      q, p, vba := Node_dest(q);
      r, m := q.(Node[Prod[a, Color]]);
      if m {
        vca, r, vea := Node_dest(r);
        _ = r;
        vfa, d := Pair_dest(r);
        if d == (Color(Black{})) {
          _ = p;
          va, e := Pair_dest(p);
          if e == (Color(Black{})) {
            ab := c;
            t2a := t4;
            return Tree[Prod[a, Color]](Node[Prod[a, Color]]{Tree[Prod[a, Color]](Node[Prod[a, Color]]{Tree[Prod[a, Color]](Node[Prod[a, Color]]{vca, Prod[a, Color]{vfa, Color(Black{})}, vea}), Prod[a, Color]{va, Color(Black{})}, vba}), Prod[a, Color]{ab, Color(Black{})}, t2a});
          }
        }
      }
    }
  };
  {
    q, m := x0.(Node[Prod[a, Color]]);
    if m {
      p, vaa, d := Node_dest(q);
      if d == (Tree[Prod[a, Color]](Leaf[Prod[a, Color]]{})) {
        q, m := p.(Node[Prod[a, Color]]);
        if m {
          vca, q, vea := Node_dest(q);
          _ = q;
          vfa, e := Pair_dest(q);
          if e == (Color(Black{})) {
            ab := c;
            t2a := t4;
            return Tree[Prod[a, Color]](Node[Prod[a, Color]]{Tree[Prod[a, Color]](Node[Prod[a, Color]]{Tree[Prod[a, Color]](Node[Prod[a, Color]]{vca, Prod[a, Color]{vfa, Color(Black{})}, vea}), vaa, Tree[Prod[a, Color]](Leaf[Prod[a, Color]]{})}), Prod[a, Color]{ab, Color(Black{})}, t2a});
          }
        }
      }
    }
  };
  {
    q, m := x0.(Node[Prod[a, Color]]);
    if m {
      q, vaa, p := Node_dest(q);
      r, m := q.(Node[Prod[a, Color]]);
      if m {
        vca, r, vea := Node_dest(r);
        _ = r;
        vfa, d := Pair_dest(r);
        if d == (Color(Black{})) {
          s, m := p.(Node[Prod[a, Color]]);
          if m {
            vb, s, vga := Node_dest(s);
            _ = s;
            vha, e := Pair_dest(s);
            if e == (Color(Black{})) {
              ab := c;
              t2a := t4;
              return Tree[Prod[a, Color]](Node[Prod[a, Color]]{Tree[Prod[a, Color]](Node[Prod[a, Color]]{Tree[Prod[a, Color]](Node[Prod[a, Color]]{vca, Prod[a, Color]{vfa, Color(Black{})}, vea}), vaa, Tree[Prod[a, Color]](Node[Prod[a, Color]]{vb, Prod[a, Color]{vha, Color(Black{})}, vga})}), Prod[a, Color]{ab, Color(Black{})}, t2a});
            }
          }
        }
      }
    }
  };
  {
    q, m := x0.(Node[Prod[a, Color]]);
    if m {
      va, p, vba := Node_dest(q);
      _ = p;
      vca, d := Pair_dest(p);
      if d == (Color(Black{})) {
        ab := c;
        t2a := t4;
        return Tree[Prod[a, Color]](Node[Prod[a, Color]]{Tree[Prod[a, Color]](Node[Prod[a, Color]]{va, Prod[a, Color]{vca, Color(Black{})}, vba}), Prod[a, Color]{ab, Color(Black{})}, t2a});
      }
    }
  };
  panic("match failed");
}
\end{lstlisting}

\end{document}